\documentclass[prb,showpacs,twocolumn]{revtex4}%
\usepackage{amsfonts}
\usepackage{amsmath}
\usepackage{amssymb}
\usepackage{graphicx}
\usepackage{times}%
\providecommand{\U}[1]{\protect\rule{.1in}{.1in}}
\begin{document}
\title{Persistent spin helix in Rashba-Dresselhaus two-dimensional electron systems}
\author{Ming-Hao\ Liu}
\affiliation{Department of Physics, National Taiwan University, Taipei 10617, Taiwan}
\author{Kuo-Wei Chen}
\affiliation{Department of Physics, National Taiwan University, Taipei 10617, Taiwan}
\author{Son-Hsien Chen}
\affiliation{Department of Physics, National Taiwan University, Taipei 10617, Taiwan}
\author{Ching-Ray Chang}
\affiliation{Department of Physics, National Taiwan University, Taipei 10617, Taiwan}


\begin{abstract}
A persistent spin helix (PSH) in spin-orbit-coupled two-dimensional electron
systems was recently predicted to exist in two cases: [001] quantum wells
(QWs) with equal coupling strengths of the Rashba and the Dresselhaus
interactions (RD), and Dresselhaus-only [110] QWs. Here we present supporting
results and further investigations, using our previous results [Phys. Rev. B
\textbf{72}, 153305 (2005)]. Refined PSH patterns for both RD [001] and
Dresselhaus [110] QWs are shown, such that the feature of the helix is clearly
seen. We also discuss the time dependence of spin to reexamine the origin of
the predicted persistence of the PSH. For the RD [001] case, we further take
into account the random Rashba effect, which is much more realistic than the
constant Rashba model. The distorted PSH pattern thus obtained suggests that
such a PSH may be more observable in the Dresselhaus [110] QWs, if the dopants
cannot be regularly enough distributed.

\end{abstract}
\date{\today}
\maketitle

\section{Introduction}

The spin precession effect due to spin-orbit coupling in low-dimensional
semiconductor structures,\cite{Winkler,MHL,MHL2} has been one of the most
important and attractive topics in the emerging field of
spintronics,\cite{RMP: spintronics} for it is not only a beautiful physical
phenomenon but also plays the central role in spintronic device proposals,
such as the spin-field-effect transistor.\cite{Datta-Das} Whereas the
spin-orbit interaction is often $k$-dependent ($k$ is the wave vector of the
transport carrier), unavoidable momentum scattering randomizing the carrier
spin direction leads to spin relaxation. Compared to electrical spin
detection, the optical pump-probe technique has been a more successful and
reliable method to observe not only the spin precession but also the spin
relaxation in semiconductor structures.\cite{SMspintronics} However, the
success of the optical method lies only in the time-resolved local spin precession.

Due to the limit of the laser spot size, space-resolved spin precession, which
is within the length scales of e.g., 0.1 $%
%TCIMACRO{\unit{\U{3bc}m}}%
%BeginExpansion
\operatorname{\mu m}%
%EndExpansion
$ for InAs-based quantum wells (QWs) and 1 $%
%TCIMACRO{\unit{\U{3bc}m}}%
%BeginExpansion
\operatorname{\mu m}%
%EndExpansion
$ for GaAs-based QWs, is so far only theoretically understood. In
two-dimensional electron systems (2DESs), neither the predicted Rashba nor
Dresselhaus spin precession patterns\cite{MHL,MHL2} have been experimentally
observed. Fundamental difficulties in experimentally observing the
space-resolved spin precession may lie in the need for either higher
resolution (smaller laser spots) or suppression of the spin relaxation.
Schliemann \textit{et al.} previously suggested that in [001] QWs the $k$
dependence of the spin-orbit fields can be removed when reaching the condition
where the Rashba equals the Dresselhaus coupling strengths
(RD).\cite{Schliemann} It turns out that the D'yakonov-Perel' spin
relaxation\cite{DP} in 2DESs may be greatly suppressed under this RD
condition. Another interesting case is a Dresselhaus [110] QW (i.e., growth
direction along [110] with only the Dresselhaus interaction), where the
spin-orbit field directions are also constant in $k$. Recent experiments
indeed revealed evidence supporting the uniqueness of slower spin relaxation
rates.\cite{110experiments}

Very recently, Bernevig \textit{et al.} claimed that the above-mentioned spin
precession phenomena should be experimentally observable in certain
geometries.\cite{PSH} They found an exact $SU\left(  2\right)  $ symmetry in
2DESs for both the RD [001] and the Dresselhaus [110] models. In these two
cases the special form of the Hamiltonian leads to $\left[  S_{i}%
,\mathcal{H}\right]  =0$, where $S_{i}$ is the $i$ component of the spin
operator, such that the spin does not depend on time (infinite spin lifetime).
They also predicted a persistent spin helix (PSH), which is a special spin
precession pattern with the precession angle depending only on the net
displacement in certain directions ($\pm\lbrack110]$ for the RD [001] model
and $\pm\lbrack1\bar{1}0]$ for the Dresselhaus [110] model).

In this paper, we first demonstrate that our previous work,\cite{MHL} where an
analytical formula describing the spin vector as a function of position for an
injected spin was presented, exactly implied such a PSH in the RD [001] case.
We next use the same method to obtain the spin vector formula for the
Dresselhaus [110] model. For both cases, we present refined PSH patterns using
our formulas and considering InAs-based 2DESs. Whereas the Rashba field partly
stems from the ionized dopants in the vicinity of the 2DES layer and will
never be a constant in reality, we will also show the influence due to the
random Rashba effect\cite{Sherman} on the PSH pattern in RD [001] QWs.

This paper is organized as follows. We introduce and derive the PSH in Sec.
\ref{sec PSH}, where we review our basic formalism, examine the PSH in RD
[001] and Dresselhaus [110] QWs using the constant Rashba model, reexamine the
time dependence of the spin in such systems, and show the distortion effect
induced by the random Rashba field for the RD [001] case. Finally, we conclude
in Sec. \ref{sec Conclusion}. Throughout this paper, single-particle quantum
mechanics is applied, the clean limit of the 2DES is assumed, and the
effective mass approximation in mesoscopic semiconductor structures is adopted.

\section{Persistent Spin Helix\label{sec PSH}}

In this section we first introduce the basic formalism of calculating the
space-resolved spin vectors, and then apply the formalism to some special
cases of spin-orbit-coupled 2DESs, namely, the RD [001] and the Dresselhaus
[110] QWs, where PSH is predicted to exist. In order to examine the origin of
the persistency of the PSH, we will also reexamine the time evolution of the
spin vectors. Finally, how seriously the random Rashba effect\cite{Sherman}
will damage the PSH in the RD [001] case will be shown.

\subsection{Basic formalism\label{SecFormalism}}

\begin{figure}
[ptb]
\begin{center}
\includegraphics[
height=1.7175in,
width=3.1661in
]%
{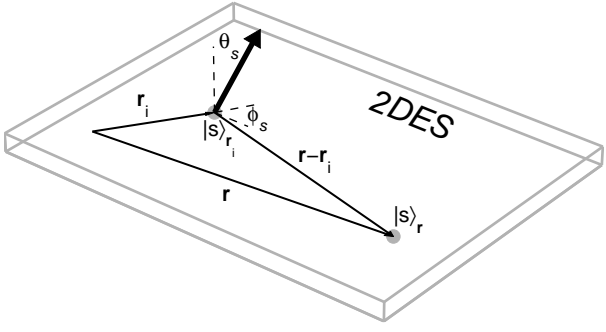}%
\caption{Schematic sketch of a 2DES, where an electron spin described by a
given state ket $\left\vert s\right\rangle _{\vec{r}_{i}}$ is ideally injected
at $\vec{r}_{i}$ with polar and azimuthal angles $\theta_{s}$ and $\phi_{s}$,
respectively.}%
\label{SCHfig}%
\end{center}
\end{figure}

Suppose that an electron spin is ideally injected at $\vec{r}_{i}$ inside the
2DES, and is described by a given state ket $\left\vert s\right\rangle
_{\vec{r}_{i}}$, where the label $s$ includes both space and spin information
of the electron. See Fig. \ref{SCHfig}. The basic strategy to obtain the
space-resolved spin vectors for the injected spin is to first obtain the state
ket $\left\vert s\right\rangle _{\vec{r}}$ on the arbitrary position $\vec
{r}=\left(  x,y\right)  $, subject to the given ket $\left\vert s\right\rangle
_{\vec{r}_{i}}$. Once $\left\vert s\right\rangle _{\vec{r}}$ is obtained, we
can determine the most probable direction, in which the spin will be pointing
at that place $\vec{r}$, by calculating the quantum expectation values
$\left\langle \vec{\sigma}\right\rangle _{\vec{r}}=$ $_{\vec{r}}\left\langle
s\right\vert \vec{\sigma}\left\vert s\right\rangle _{\vec{r}}$, where
$\vec{\sigma}=\left(  \sigma_{x},\sigma_{y},\sigma_{z}\right)  $ is the Pauli
matrix vector. The spin vector as a function of $\vec{r}$ is then given by
$\langle\vec{S}\rangle_{\vec{r}}\equiv\left(  \hbar/2\right)  \left\langle
\vec{\sigma}\right\rangle _{\vec{r}}$.%
%TCIMACRO{\FRAME{ftbpFU}{3.1661in}{1.7175in}{0pt}{\Qcb{Schematic sketch of a
%2DES, where an electron spin described by a given state ket $\left\vert
%s\right\rangle _{\vec{r}_{i}}$ is ideally injected at $\vec{r}_{i}$ with polar
%and azimuthal angles $\theta_{s}$ and $\phi_{s}$, respectively.}}%
%{\Qlb{SCHfig}}{fig1.ps}{\special{ language "Scientific Word";
%type "GRAPHIC";  maintain-aspect-ratio TRUE;  display "ICON";
%valid_file "F";  width 3.1661in;  height 1.7175in;  depth 0pt;
%original-width 6.2759in;  original-height 3.3788in;  cropleft "0";
%filename 'FIG1.ps';file-properties "XNPEU";}}}%
%BeginExpansion
%EndExpansion

Specifically, how the given state ket $\left\vert s\right\rangle _{\vec{r}%
_{i}}$ after being spatially translated from $\vec{r}_{i}$ to $\vec{r}$ is
simply obtained by operating on it with the translation operator
$\mathcal{T}\left(  \vec{a}\right)  \equiv\exp\left(  -i\vec{p}\cdot\vec
{a}/\hbar\right)  $: $\left\vert s\right\rangle _{\vec{r}}=\left\vert
s\right\rangle _{\vec{r}_{i}\rightarrow\vec{r}}=\mathcal{T}\left(  \vec{r}%
_{i}-\vec{r}\right)  \left\vert s\right\rangle _{\vec{r}_{i}}$. Note that here
we have assumed ideal point injection of spin. If the spins are injected via a
finite-size contact, we have to sum over all the states coming from all the
possible injection points:\cite{MHL2} $\left\vert s\right\rangle _{\vec{r}%
}\propto\sum_{i}\left\vert s\right\rangle _{\vec{r}_{i}\rightarrow\vec{r}}$.
In general, for spin-orbit-coupled 2DESs we can obtain two spin-dependent
eigenstates $|\sigma,\vec{k}_{\sigma}\rangle$, where $\sigma=\uparrow$ or
$\downarrow$ is the index of the spin state subject to the wave vector
$\vec{k}_{\sigma}$, and then expand the injected state ket using this basis:
$\left\vert s\right\rangle _{\vec{r}_{i}}=\sum_{\sigma=\uparrow,\downarrow
}|\sigma,\vec{k}_{\sigma}\rangle\langle\sigma,\vec{k}_{\sigma}|s\rangle
_{\vec{r}_{i}}$. Operation of $\exp\left(  -i\vec{p}\cdot\vec{a}/\hbar\right)
$ on $\left\vert s\right\rangle _{\vec{r}_{i}}$ immediately gives different
phases to the two components since $\exp\left(  -i\vec{p}\cdot\vec{a}%
/\hbar\right)  |\sigma,\vec{k}_{\sigma}\rangle=\exp(-i\vec{k}_{\sigma}%
\cdot\vec{a})|\sigma,\vec{k}_{\sigma}\rangle$ with $\left\vert k_{\uparrow
}\right\vert \neq\left\vert k_{\downarrow}\right\vert $ in general. Such a
phase difference, in fact, plays a key role in the spatial spin precession.

Next we will use the above strategy to solve two concrete examples: the RD
[001] and the Dresselhaus [110] QWs.\vspace{0.3in}

\subsection{PSH in RD [001] QWs}

In this case we consider a [001]-grown 2DES confined in an asymmetric QW
(where the Rashba effect\cite{Rashba term} exists) made of III-V
semiconductors (where the Dresselhaus coupling\cite{Dresselhaus term} exists).
Using the linear Rashba and the linear Dresselhaus model, the Hamiltonian can
be written as%
\begin{equation}
\mathcal{H}_{\left[  001\right]  }=\frac{p_{x}^{2}+p_{y}^{2}}{2m^{\star}%
}+\frac{\alpha}{\hbar}\left(  p_{y}\sigma_{x}-p_{x}\sigma_{y}\right)
+\frac{\beta}{\hbar}\left(  p_{x}\sigma_{x}-p_{y}\sigma_{y}\right)  ,
\label{H_001}%
\end{equation}
where $\alpha$ and $\beta$, assumed to be constant at present, are the Rashba
and Dresselhaus strengths, respectively, and $m^{\star}$ is the electron
effective mass. The well-known eigenfunctions are%
\begin{equation}
\langle\vec{r}|\uparrow\downarrow,\vec{k}_{\uparrow\downarrow}\rangle
=\frac{e^{i\vec{k}_{\uparrow\downarrow}\cdot\vec{r}}}{\sqrt{2}}\left(
\begin{array}
[c]{c}%
ie^{-i\varphi}\\
\pm1
\end{array}
\right)  , \label{RDeigen}%
\end{equation}
corresponding to the eigenenergies%
\begin{equation}
E_{\uparrow\downarrow}(\vec{k})=\frac{\hbar^{2}k^{2}}{2m^{\star}}\pm
\zeta\left(  \alpha,\beta,\phi\right)  k. \label{RDeigenE}%
\end{equation}
Here we have defined%
\begin{equation}
\varphi\equiv\arg\left[  \alpha\cos\phi+\beta\sin\phi+i(\alpha\sin\phi
+\beta\cos\phi)\right]  \label{varphi}%
\end{equation}
and%
\begin{equation}
\zeta\left(  \alpha,\beta,\phi\right)  =\sqrt{\alpha^{2}+\beta^{2}%
+2\alpha\beta\sin2\phi}, \label{zeta}%
\end{equation}
where $\phi$ is the argument angle of $\vec{k}\equiv\left(  k_{x}%
,k_{y}\right)  =k\left(  \cos\phi,\sin\phi\right)  $. Using the formalism
introduced in Sec. \ref{SecFormalism} and assuming that the spin is injected
at $\vec{r}_{i}=\left(  0,0\right)  $, the spin vector formula, which had been
previously derived,\cite{MHL} reads\begin{widetext}%
\begin{equation}
\left\langle \vec{\sigma}\right\rangle _{\vec{r}}^{\text{001}}=\left(
\begin{array}
[c]{c}%
-\cos\theta_{s}\cos\varphi\sin\Delta\theta_{\vec{r}}+\sin\theta_{s}\left(
\cos\phi_{s}\cos^{2}\dfrac{\Delta\theta_{\vec{r}}}{2}-\cos\left(
2\varphi-\phi_{s}\right)  \sin^{2}\dfrac{\Delta\theta_{\vec{r}}}{2}\right)  \\
-\cos\theta_{s}\sin\varphi\sin\Delta\theta_{\vec{r}}+\sin\theta_{s}\left(
\sin\phi_{s}\cos^{2}\dfrac{\Delta\theta_{\vec{r}}}{2}-\sin\left(
2\varphi-\phi_{s}\right)  \sin^{2}\dfrac{\Delta\theta_{\vec{r}}}{2}\right)  \\
\cos\theta_{s}\cos\Delta\theta_{\vec{r}}+\sin\theta_{s}\cos\left(
\varphi-\phi_{s}\right)  \sin\Delta\theta_{\vec{r}}%
\end{array}
\right)  ,\label{<S>001}%
\end{equation}
where\end{widetext}%
\begin{equation}
\Delta\theta_{\vec{r}}\equiv\frac{2m^{\star}\zeta\left(  \alpha,\beta
,\phi\right)  r}{\hbar^{2}} \label{Dtheta}%
\end{equation}
is the spin precession angle. Note that the position dependence of Eq.
(\ref{<S>001}) lies in $\varphi$ and $\Delta\theta_{\vec{r}}$. Now we apply
the condition $\alpha=\beta$ to fit the ReD condition. Obviously, we obtain
$\varphi\left(  \alpha=\beta\right)  =\pi/4$ and $\zeta\left(  \alpha
,\alpha,\phi\right)  =2\alpha\cos\left(  \pi/4-\phi\right)  $, leading to the
precession angle
\begin{equation}
\Delta\theta_{\vec{r}}=\frac{4m^{\star}\alpha}{\hbar^{2}}r_{110}\equiv
\Delta\theta_{\text{RD}} \label{thetaReD}%
\end{equation}
with $r_{110}\equiv r\cos\left(  \pi/4-\phi\right)  $. It turns out that the
position dependence of the spin vector is now reduced to only $\Delta
\theta_{\vec{r}}$ through Eq. (\ref{thetaReD}), which depends only on the net
displacement in the [110] direction. Thus our previous results had indeed
implied the PSH proposed by Bernevig \textit{et al.}\cite{PSH}%
%TCIMACRO{\FRAME{ftbpFU}{3.2638in}{2.9525in}{0pt}{\Qcb{(Color online) Effective
%magnetic fields drawn in the $k_{x}$-$k_{y}$ coordinates for (a) the RD model,
%and (c) the Dresselhaus [110] model. (b) and (d) show the corresponding PSH
%patterns in a $1\times1$ $\unit{\U{3bc}m}^{2}$ InAs-based 2DES, with assumed
%point injection of spins indicated by the bold arrows. Color bar calibrates
%$\left\langle S_{z}\right\rangle $ in (b) and $\left\langle S_{x}\right\rangle
%$ in (d), in units of $\hbar/2$, with yellow (bright) and red (dark) colors
%meaning positive and negative values, respectively.}}{\Qlb{PSHfig}}%
%{fig2.ps}{\special{ language "Scientific Word";  type "GRAPHIC";
%maintain-aspect-ratio TRUE;  display "ICON";  valid_file "F";
%width 3.2638in;  height 2.9525in;  depth 0pt;  original-width 6.2906in;
%original-height 5.4379in;  cropleft "0";  croptop "1";  cropright "1";
%cropbottom "0";  filename 'FIG2.ps';file-properties "XNPEU";}}}%
%BeginExpansion
\begin{figure}
[ptb]
\begin{center}
\includegraphics[
height=2.9525in,
width=3.2638in
]%
{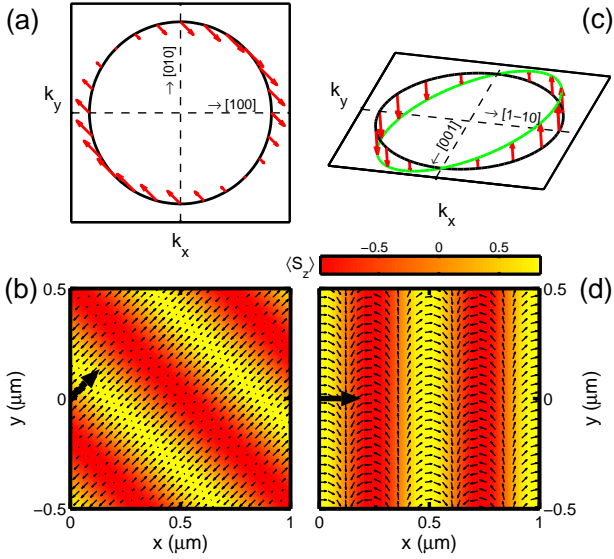}%
\caption{(Color online) Effective magnetic fields drawn in the $k_{x}$-$k_{y}$
coordinates for (a) the RD model, and (c) the Dresselhaus [110] model. (b) and
(d) show the corresponding PSH patterns in a $1\times1$ $\operatorname{\mu
m}^{2}$ InAs-based 2DES, with assumed point injection of spins indicated by
the bold arrows. Color bar calibrates $\left\langle S_{z}\right\rangle $ in
(b) and $\left\langle S_{x}\right\rangle $ in (d), in units of $\hbar/2$, with
yellow (bright) and red (dark) colors meaning positive and negative values,
respectively.}%
\label{PSHfig}%
\end{center}
\end{figure}
%EndExpansion

In fact, in the RD condition the effective magnetic field generated by the
Rashba and Dresselhaus spin-orbit couplings can be depicted in the momentum
space as shown in Fig. \ref{PSHfig}(a), where the field directions are
constant (pointing to either [1\={1}0] or [\={1}10]) while the field strength
depends on the projection of $k$ on [110] direction. This can be seen by
defining the spin-orbit (effective magnetic) field as $\vec{B}_{\text{eff}%
}\equiv\zeta\langle\uparrow,\vec{k}_{\uparrow}|\vec{\sigma}|\uparrow,\vec
{k}_{\uparrow}\rangle$, so that with the RD condition we have%
\begin{equation}
\vec{B}_{\text{RD}}=\sqrt{2}\alpha\cos\left(  \dfrac{\pi}{4}-\phi\right)
\left(  1,-1,0\right)  \text{ .} \label{BReD}%
\end{equation}

Now, we consider a $1\times1$ $%
%TCIMACRO{\unit{\U{3bc}m}}%
%BeginExpansion
\operatorname{\mu m}%
%EndExpansion
^{2}$ InAs-based 2DES and apply Eq. (\ref{<S>001}) with $\alpha=\beta$ to
obtain the PSH. Inside the 2DES, $m^{\star}=0.023m_{0}$, $m_{0}$ being the
electron rest mass, and the Dresselhaus coupling strength is put at
$\beta=1.062\times10^{-2}$ $%
%TCIMACRO{\unit{eV}}%
%BeginExpansion
\operatorname{eV}%
%EndExpansion%
%TCIMACRO{\unit{nm}}%
%BeginExpansion
\operatorname{nm}%
%EndExpansion
$ to simulate a 5-$%
%TCIMACRO{\unit{nm}}%
%BeginExpansion
\operatorname{nm}%
%EndExpansion
$-thick InAs quantum well.\cite{Knap} As shown in Fig. \ref{PSHfig}(b), we
have obtained the PSH in agreement with the prediction of Bernevig \textit{et
al}., in a more concrete way. Note that here we have assumed a point spin
injection at $\left(  x=0,y=0\right)  $ with spin polarization parallel to [110].

\subsection{PSH in Dresselhaus [110] QWs}

Next we consider the Dresselhaus [110] model, where the 2DES is grown along
[110] and is described by the Hamiltonian%
\begin{equation}
\mathcal{H}_{\left[  110\right]  }=\frac{p_{x}^{2}+p_{y}^{2}}{2m}-2\frac
{\beta}{\hbar}p_{x}\sigma_{z}, \label{H_110}%
\end{equation}
which is in a diagonal form. The eigenfuntions are
\begin{equation}
\left\langle \vec{r}|\uparrow,k_{\uparrow}\right\rangle =e^{i\vec{k}%
_{\uparrow}\cdot\vec{r}}\left\vert \uparrow\right\rangle \doteq e^{i\vec
{k}_{\uparrow}\cdot\vec{r}}\left(
\begin{array}
[c]{c}%
1\\
0
\end{array}
\right)
\end{equation}
and%
\begin{equation}
\left\langle \vec{r}|\downarrow,k_{\downarrow}\right\rangle =e^{i\vec
{k}_{\downarrow}\cdot\vec{r}}\left\vert \downarrow\right\rangle \doteq
e^{i\vec{k}_{\downarrow}\cdot\vec{r}}\left(
\begin{array}
[c]{c}%
0\\
1
\end{array}
\right)  ,
\end{equation}
corresponding to the eigenenergies%
\begin{equation}
E_{\uparrow\downarrow}(\vec{k})=\frac{\hbar^{2}k^{2}}{2m^{\star}}\mp2\beta
k_{x}. \label{DeigenE}%
\end{equation}
Using the formalism introduced in Sec. \ref{SecFormalism}, we obtain the
spatially evolved state ket%
\begin{equation}
\left\vert s\right\rangle _{\vec{r}_{i}\rightarrow\vec{r}}=\sum_{\sigma
=\uparrow,\downarrow}\exp\left(  -\frac{i\Delta\theta_{\text{110}}}{2}\right)
c_{\sigma}\left\vert \sigma\right\rangle , \label{sket110}%
\end{equation}
with the phase%
\begin{equation}
\Delta\theta_{\text{110}}=\left(  k_{\downarrow}-k_{\uparrow}\right)
r=\frac{4m^{\star}\beta}{\hbar^{2}}r_{\text{1\={1}0}}, \label{theta110}%
\end{equation}
where $c_{\sigma}=\left\langle \sigma|s\right\rangle _{\vec{r}_{i}}$ are the
expansion coefficients and $r_{\text{1\={1}0}}\equiv r\cos\phi$ is the
displacement along the $\phi=0$ direction. Note that in this Dresselhaus [110]
model, the $x$ direction ($\phi=0$) is parallel with [1\={1}0]. As in the RD
case, the precession angle depends on the net displacement along [1\={1}0].

Applying Eq. (\ref{sket110}) with expansion coefficients $c_{\uparrow
}=e^{-i\phi_{s}}\cos\left(  \theta_{s}/2\right)  $ and\ $c_{\downarrow}%
=\sin\left(  \theta_{s}/2\right)  $, we obtain the spin vector formula for the
Dresselhaus [110] model:%
\begin{equation}
\left\langle \vec{\sigma}\right\rangle _{\vec{r}}^{\text{110}}=\left(
\begin{array}
[c]{c}%
\sin\theta_{s}\left(  \cos\phi_{s}\cos\Delta\theta_{\text{110}}-\sin\phi
_{s}\sin\Delta\theta_{\text{110}}\right) \\
-\sin\theta_{s}\left(  \cos\phi_{s}\sin\Delta\theta_{\text{110}}+\sin\phi
_{s}\cos\Delta\theta_{\text{110}}\right) \\
\cos\theta_{s}%
\end{array}
\right)  . \label{<S>110}%
\end{equation}
Comparing Eqs. (\ref{DeigenE}) with (\ref{RDeigenE}), the spin splitting
linear in $k$ in this Dresselhaus [110] case is $2\beta k_{x}=\left(
2\beta\cos\phi\right)  k$, so we can define%
\begin{equation}
\vec{B}_{\text{110}}\equiv2\beta\cos\phi\left\langle \uparrow\right\vert
\vec{\sigma}\left\vert \uparrow\right\rangle =2\beta k\cos\phi\left(
0,0,1\right)  \label{B110}%
\end{equation}
to depict the effective magnetic field in the $k$ space, as shown in Fig.
\ref{PSHfig}(c). Correspondingly, the PSH pattern is drawn in Fig.
\ref{PSHfig}(d), where the same material parameters (including the Dresselhaus
strength) as in Fig. \ref{PSHfig}(b) are taken (except that there is no
$\alpha$ here), and an $x$-polarized spin is injected at $\left(  0,0\right)
$.

\subsection{More on the spin precession: Space translation vs. time evolution}

\begin{figure}
[ptb]
\begin{center}
\includegraphics[
height=2.7017in,
width=3.2128in
]%
{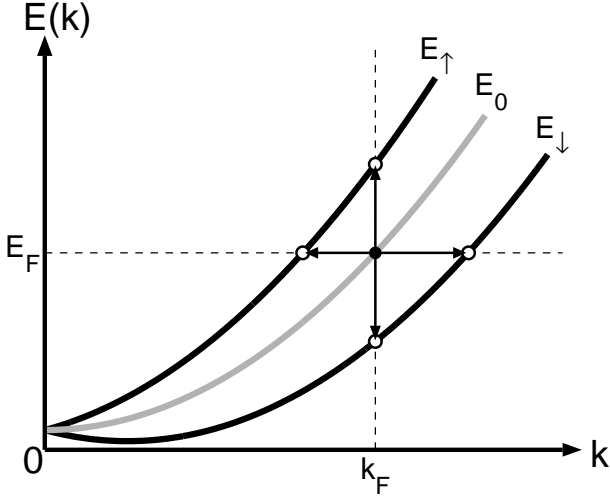}%
\caption{Band structure of a spin-orbit coupled 2DES. Vertical projection
fixes $k$ as $k_{F}$, leading to time-resolved spin precession, while
horizontal projection fixes energy as $E_{F}$, leading to spatial spin
precession.}%
\label{PROJfig}%
\end{center}
\end{figure}

Before we move on to the influence on the PSH due to the random Rashba effect,
below we provide a series of discussion over the time evolution of spin in
these spin-orbit coupled systems. So far, the above position-dependent spin
vectors are based on the time-independent Schr\"{o}dinger approach. In fact,
the key to such space-resolved spin precession lies in the assumption of
$E_{\uparrow}=E_{\downarrow}=E_{F}$, $E_{F}$ being the Fermi energy. Under
this assumption, the injected spin is projected onto two available states at
the Fermi level with equal energy but different momentums (see the
\textquotedblleft horizontal projection\textquotedblright\ in Fig.
\ref{PROJfig}). The difference in the projected wave vectors, as we mentioned
in Sec. \ref{SecFormalism}, leads to a phase difference between the two spin
components of the total wave function when spatially translated by a distance,
and an angle of rotation of the spin then occurs. Meanwhile, with
$E_{\uparrow}=E_{\downarrow}$ the wave function does not have nontrivial time
dependence since the time evolution operator $\exp\left(  -i\mathcal{H}%
t/\hbar\right)  $ produces equal phases for both spin components. In other
words, the persistence for the spin precession pattern, or the PSH, is merely
a built-in property, once the projected states satisfy $E_{\uparrow
}=E_{\downarrow}$.%
%TCIMACRO{\FRAME{ftbpFU}{3.2128in}{2.7017in}{0pt}{\Qcb{Band structure of a
%spin-orbit coupled 2DES. Vertical projection fixes $k$ as $k_{F}$, leading to
%time-resolved spin precession, while horizontal projection fixes energy as
%$E_{F}$, leading to spatial spin precession.}}{\Qlb{PROJfig}}{fig3.ps}%
%maintain-aspect-ratio TRUE;  display "ICON";  valid_file "F";
%width 3.2128in;  height 2.7017in;  depth 0pt;  original-width 7.1252in;
%original-height 5.7597in;  cropleft "0";  croptop "1";  cropright "1";
%cropbottom "0";  filename 'FIG3.ps';file-properties "XNPEU";}}}%
%BeginExpansion
%EndExpansion

From the quantum-mechanical viewpoint, it is only when the injected electron
is projected onto states with $E_{\uparrow}\neq E_{\downarrow}$ that leads to
nontrivial time dependence. If we regard the spin-orbit interaction as an
effective magnetic field $\vec{B}_{\text{eff}}$ and treat the injected
electron spin as a constantly moving particle with wave vector $k_{F}$ in the
2DES, the sudden turn-on of $\vec{B}_{\text{eff}}$ splits the energy level the
electron originally occupies, leaving the momentum it carries unchanged. See
the \textquotedblleft vertical projection\textquotedblright\ in Fig.
\ref{PROJfig}. Taking the Rashba-Dresselhaus [001] 2DES for example (not
necessarily RD), the split levels are indicated in Eq. (\ref{RDeigenE}), or
simply $E_{0}\pm\Delta E$ with $\Delta E=\zeta\left(  \alpha,\beta
,\phi\right)  k$ and $E_{0}=\hbar^{2}k^{2}/2m^{\star}$. It turns out that
different phases $\exp\left(  \mp i\Delta Et/\hbar\right)  $ for the two spin
components will emerge, under the operation of $\exp\left(  -i\mathcal{H}%
t/\hbar\right)  $. Thus we can analogously use the ket $\left\vert
s\right\rangle _{t}=e^{iE_{0}t/\hbar}(c_{\uparrow}e^{-i\Delta\theta_{t}%
/2}|\uparrow,\vec{k}_{\uparrow}\rangle+c_{\downarrow}e^{i\Delta\theta_{t}%
/2}|\downarrow,\vec{k}_{\downarrow}\rangle)$ to derive the time-resolved
spin-vector formulas, which are exactly of the same forms with the spatial
ones, except that $\Delta\theta_{\vec{r}}$ is replaced by $\Delta\theta
_{t}=2\zeta kt/\hbar$.

Quite interestingly, both methods (space translation and time evolution) lead
to exactly the same precession angle, if we treat the electron as a moving
particle with group velocity $v_{g}=\hbar k/m^{\star}$. For example, in the
spatial treatment for the Rashba-Dresselhaus [001] QWs, the spin precession
angle between two points $L$ apart is $\Delta\theta_{\vec{r}}=2m^{\star}\zeta
L/\hbar^{2}$, while in the time treatment we have $\Delta\theta_{t}=2\zeta
kL/v_{g}\hbar$, such that $\Delta\theta_{\vec{r}}=\Delta\theta_{t}$ is
guaranteed. For a fixed propagation direction, the precession angle between
any pair of injection-detection points depends only on the distance, but not
the velocity (or the wave vector) of the electron. This is similar to a
rolling disk without going into a slide in classical physics. In fact, these
two ideal assumptions (horizontal and vertical projections shown in Fig.
\ref{PROJfig}) are often adopted, although in real situations combined effects
are more likely to occur. The actual projection of states may depend on the
temperature fluctuation, interaction between the injected electrons and the
2DES, and even the applied bias. These may require further studies but we end
the discussion here.

Finally, it is of much pedagogical meaning to relate the Schr\"{o}dinger and
the Heisenberg formalisms for the time evolution of spin in the
Rashba-Dresselhaus problems. If we fix the electron momentum as $\hbar k$ (the
vertical projection), from the Hamiltonian for, e.g., the Rashba-Dresselhaus
[001] general case shown in Eq. (\ref{H_001}) we can solve the time dependence
of the spin operators in the Heisenberg picture. Note that here we treat
$p=\hbar k$ in Eq. (\ref{H_001}) simply as numbers. Only the $\sigma$'s
therein are operators. Using the Heisenberg equation of motion $i\hbar
\dot{\sigma}_{i}=\left[  \sigma_{i},\mathcal{H}\right]  $, one can obtain the
differential equations governing the time evolution of the spin components:
$\dot{\sigma}_{x}=a\sigma_{z},$ $\dot{\sigma}_{y}=b\sigma_{z},$ and
$\dot{\sigma}_{z}=-a\sigma_{x}-b\sigma_{y}$, with $a\equiv2\left(  \alpha
k_{x}+\beta k_{y}\right)  /\hbar$ and $b\equiv2\left(  \alpha k_{y}+\beta
k_{x}\right)  /\hbar$. These differential equations can be easily solved to
obtain%
\begin{subequations}
\begin{align}
\left\langle \sigma_{x}\right\rangle _{t}  &  =\left[  \left\langle \sigma
_{z}\right\rangle _{0}\sin\omega t-\left(  \left\langle \sigma_{x}%
\right\rangle _{0}\cos\varphi+\sigma_{y}^{0}\sin\varphi\right)  \right.
\nonumber\\
&  \times\left.  \left(  1-\cos\omega t\right)  \right]  \cos\varphi
+\left\langle \sigma_{x}\right\rangle _{0}\label{sx_H}\\
\left\langle \sigma_{y}\right\rangle _{t}  &  =\left[  \left\langle \sigma
_{z}\right\rangle _{0}\sin\omega t-\left(  \left\langle \sigma_{x}%
\right\rangle _{0}\cos\varphi+\left\langle \sigma_{y}\right\rangle _{0}%
\sin\varphi\right)  \right. \nonumber\\
&  \times\left.  \left(  1-\cos\omega t\right)  \right]  \sin\varphi
+\left\langle \sigma_{y}\right\rangle _{0}\label{sy_H}\\
\left\langle \sigma_{z}\right\rangle _{t}  &  =\left\langle \sigma
_{z}\right\rangle _{0}\cos\omega t-\left(  \left\langle \sigma_{x}%
\right\rangle _{0}\cos\varphi+\left\langle \sigma_{y}\right\rangle _{0}%
\sin\varphi\right) \nonumber\\
&  \times\sin\omega t \label{sz_H}%
\end{align}
with $\omega^{2}\equiv a^{2}+b^{2}$ and $\left\langle \sigma_{i}\right\rangle
_{0}\equiv\left\langle \sigma_{i}\right\rangle _{t=0}$. Setting the initial
conditions as $\left\langle \sigma_{x}\right\rangle _{0}=\sin\theta_{s}%
\cos\phi_{s},$ $\left\langle \sigma_{y}\right\rangle _{0}=\sin\theta_{s}%
\sin\phi_{s}$, $\left\langle \sigma_{z}\right\rangle _{0}=\cos\theta_{s}$,
i.e., the injected spin with polar and azimuthal angles $\theta_{s}$ and
$\phi_{s}$, respectively (see Fig. \ref{SCHfig}), one can prove that Eqs.
(\ref{sx_H})--(\ref{sz_H}) are equivalent to Eq. (\ref{<S>001}) with
$\Delta\theta_{\vec{r}}$ replaced by $\omega t$.

\subsection{Random Rashba effect on the PSH in RD [001] QWs}

Now we go back to the PSH and focus on the RD [001] case. Whereas the Rashba
spin-orbit coupling partly stems from the Coulomb field generated from the
ionized dopants in the vicinity of the 2DES layer (usually some tens of
nanometers apart), the Rashba parameter is actually position dependent.
Usually, these dopants are randomly distributed (depending on the doping
techniques), and the resulting Rashba field may be constant in average but
fluctuating locally in nanometer scales (depending on the dopant
concentration). With this position dependence of $\alpha$, the Rashba term
previously written as $\mathcal{H}_{R}=\left(  \alpha/\hbar\right)  \left(
p_{y}\sigma_{x}-p_{x}\sigma_{y}\right)  $ in Eq. (\ref{H_001}) has to be
symmetrized as%
\end{subequations}
\begin{equation}
\mathcal{H}_{R}=\frac{1}{2\hbar}\left[  \alpha\left(  \vec{r}\right)  \left(
p_{y}\sigma_{x}-p_{x}\sigma_{y}\right)  +\left(  p_{y}\sigma_{x}-p_{x}%
\sigma_{y}\right)  \alpha\left(  \vec{r}\right)  \right]  , \label{HR}%
\end{equation}
since $\alpha\left(  \vec{r}\right)  $ in general no longer commutes with
$p_{x}$ or $p_{y}$. In this case, the full Hamiltonian is not easy to
diagonolize, and the desired eigenfunctions serving as the basis to expand the
injected spin state are not accessible. However, a convenient way of
contour-integral method can be applied to directly obtain the spatially
evolved state ket subject to the injected electron spin, without finding the
basis for the entire system.\cite{nonuniform} Thus spin vectors in such
nonuniform Rashba-Dresselhaus 2DESs turn out to be still computable with,
however, the precession angle given by a contour-integral form%
\begin{equation}
\Delta\Theta=\frac{2}{\hbar^{2}}\int_{C}m^{\star}\zeta\left(  \alpha\left(
\vec{r}\right)  ,\beta,\phi\right)  d\vec{\ell}, \label{DTheta}%
\end{equation}
where $C$ is the path the spin goes through [cf. Eq. (\ref{Dtheta}) for the
constant Rashba model]. Note that in Eq. (\ref{DTheta}), $\phi$ is the
argument angle of $d\vec{\ell}$ along the tangential direction at position
$\vec{r}$ on the path $C$, which is taken as the straight line connecting the
injection and detection points in the following numerical calculations.%
%TCIMACRO{\FRAME{ftbpFU}{3.3537in}{4.139in}{0pt}{\Qcb{(Color online) PSH
%patterns in RD [001] QWs with (a) regularly and (b) randomly distributed
%dopants. Bold arrows indicate the assumed point injection of spin. Color bar
%calibrates $\left\langle S_{z}\right\rangle $ in units of $\hbar/2$ with
%yellow (bright) and red (dark) colors meaning positive and negative values,
%respectively. Note that the $x$ axis here is pointing to [110].}%
%}{\Qlb{PSHrandfig}}{fig4.ps}{\special{ language "Scientific Word";
%type "GRAPHIC";  maintain-aspect-ratio TRUE;  display "ICON";
%valid_file "F";  width 3.3537in;  height 4.139in;  depth 0pt;
%original-width 3.2785in;  original-height 5.6835in;  cropleft "0";
%filename 'FIG4.ps';file-properties "XNPEU";}}}%
%BeginExpansion
\begin{figure}
[ptb]
\begin{center}
\includegraphics[
height=4.139in,
width=3.3537in
]%
{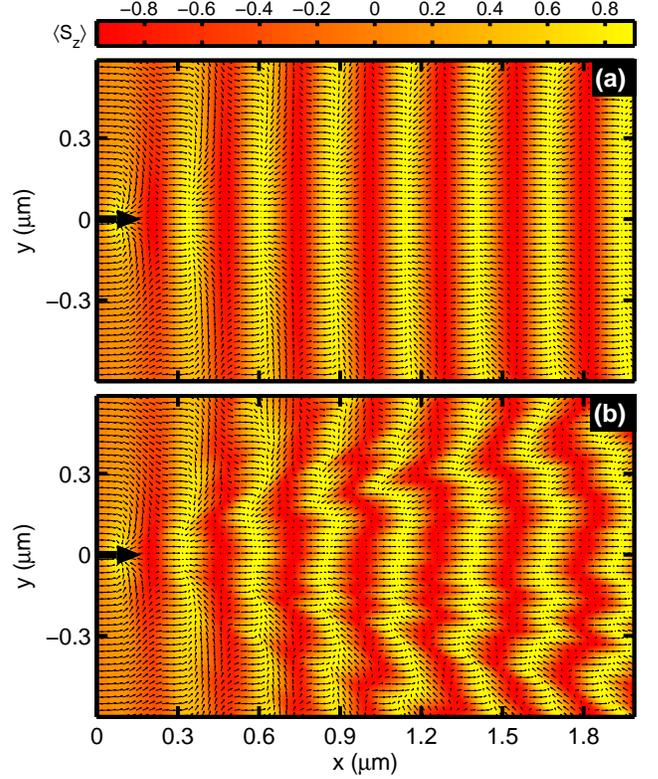}%
\caption{(Color online) PSH patterns in RD [001] QWs with (a) regularly and
(b) randomly distributed dopants. Bold arrows indicate the assumed point
injection of spin. Color bar calibrates $\left\langle S_{z}\right\rangle $ in
units of $\hbar/2$ with yellow (bright) and red (dark) colors meaning positive
and negative values, respectively. Note that the $x$ axis here is pointing to
[110].}%
\label{PSHrandfig}%
\end{center}
\end{figure}
%EndExpansion

To investigate the influence on the PSH due to such a random Rashba effect,
let us consider two $2.0\times1.2$ $%
%TCIMACRO{\unit{\U{3bc}m}}%
%BeginExpansion
\operatorname{\mu m}%
%EndExpansion
^{2}$ InAs-based 2DESs, one with regular dopant distribution and the other
with random distribution. Note that, in the latter case, the coordinate of
each dopant is completely random: probability of each dopant to occupy each
available lattice site is equally likely. Detailed process of generating the
position dependent Rashba parameter $\alpha\left(  \vec{r}\right)  $ can be
found in Refs. \onlinecite{Sherman} and \onlinecite{nonuniform}. The dopant
concentration is fixed at $\bar{n}=2.5\times10^{11}$ $%
%TCIMACRO{\unit{cm}}%
%BeginExpansion
\operatorname{cm}%
%EndExpansion
^{-2}$, leading to the average Rashba field $\left\langle \alpha\right\rangle
\approx0.019$ $%
%TCIMACRO{\unit{eV}}%
%BeginExpansion
\operatorname{eV}%
%EndExpansion%
%TCIMACRO{\unit{nm}}%
%BeginExpansion
\operatorname{nm}%
%EndExpansion
$. We put $\beta=\left\langle \alpha\right\rangle $, corresponding to the QW
thickness of around 3.7 $%
%TCIMACRO{\unit{nm}}%
%BeginExpansion
\operatorname{nm}%
%EndExpansion
$, to simulate the RD condition. Applying the contour-integral method of our
recent work,\cite{nonuniform} we plot the PSH patterns for the regular and the
random cases in Figs. \ref{PSHrandfig}(a) and \ref{PSHrandfig}(b),
respectively. Note that the plots here have been rotated so that the $x$ axis
is parallel with [110] \{instead of [100] in the previous demonstration of
Fig. \ref{PSHfig}(b)\}.

Clearly, in the regular case, even though the Rashba parameter is \emph{not
constant} in space (spatial profile may be seen in Ref.
\onlinecite{nonuniform}), the PSH pattern is still almost perfect, compared to
the one predicted in the calculation with the \emph{constant} Rashba model.
When the dopants are randomly distributed, Fig. \ref{PSHrandfig}(b) shows a
distorted PSH pattern, which may imply an intrinsic difficulty in experimental
observation of the PSH in RD [001] QWs. In addition, since the distortion
actually stems from the difference in $\Delta\Theta$ given in Eq.
(\ref{DTheta}) along different injection-detection paths, the distortion
effect may grow with distance. Indeed, one can see that the farther stripes in
Fig. \ref{PSHrandfig}(b) are more distorted than the nearer ones. Note that
here we have assumed these two somewhat wide channels to be ballistic and
unbounded, so that the distortion effect is completely induced by the random
Rashba effect.

Alternatively, the difference in $\Delta\Theta$ for different paths may
accumulate faster by raising the composite spin-orbit coupling strength
$\zeta$, or the Rashba strength $\alpha$, which increases with the dopant
concentration $\bar{n}$. When putting a higher $\bar{n}$, the stronger Rashba
strength creates a PSH pattern with more stripes, among which the farther
(nearer) ones will be more (less) distorted. In this case the distorted
pattern is similar to the one we have shown in Fig. \ref{PSHrandfig}(b),
except that only a shorter channel is required, and we do not further show here.

\section{Conclusion\label{sec Conclusion}}

To conclude, we have presented refined PSH patterns in RD [001] and the
Dresselhaus [110] QWs, first predicted by Bernevig \textit{et al.}\cite{PSH}
For the RD [001] case, we have shown that the PSH had actually been implied in
our previous results.\cite{MHL,MHL2} For the Dresselhaus [110] case, we have
also derived the spin-vector formula using the same method. For both cases, we
have shown the most important feature of the PSH: the spin precession angle
depends only on the net displacement of the injected spin along certain
directions ($\pm\lbrack110]$ for the RD [001] and $\pm\lbrack1\bar{1}0]$ for
the Dresselhaus [110] cases), and hence the name \textquotedblleft
helix.\textquotedblright\ This special feature guarantees the probability of
experimentally observing the PSH since the spin direction does not depend on
the incoming direction: momentum scattering is even allowed.

We have also discussed the time evolution of spin in the Rashba-Dresselhaus
problem. In the quantum-mechanical interpretation, the persistence of the PSH
seems to be inherent in any systems, not necessarily the RD [001] or the
Dresselhaus [110] QW, once the injected spin is projected onto states with
equal energy (the "horizontal projection" sketched in Fig. \ref{PROJfig}).
Meanwhile, when treating the electron as a constantly moving particle with the
group velocity $\hbar k/m^{\star}$, the other assumption of vertical
projection will lead to the same physics, i.e., the spin direction at a
certain position in the 2DES does not depend on time.

Applying the contour-integral method,\cite{nonuniform} we have also examined
the influence on the PSH pattern due to the random Rashba effect for the RD
[001] case. The obtained results show that unless the dopants are regularly
distributed, the PSH pattern may be distorted (though not totally destroyed)
by the random effect. In addition, the distortion effect is observed to grow
with longer channel length or higher dopant concentration (stronger Rashba
strength). Therefore, symmetric QWs grown along [110] with only the
Dresselhaus coupling may be a better candidate to observe the predicted PSH.

Finally, we remind here that in this Dresselhaus [110] model, the PSH period
given by $\pi\hbar^{2}/4m^{\star}\beta$ is quite sensitive to the electron
effective mass. For example, GaAs and InAs have similar bulk Dresselhaus
coefficients,\cite{Winkler,Knap} but a significant difference in the
two-dimensional electron effective mass.\cite{book} With $m_{\text{GaAs}%
}^{\star}/m_{\text{InAs}}^{\star}\approx3$, the PSH pattern size using GaAs
QWs will be about three times smaller than for InAs QWs, and will therefore
require experimental setups with higher resolution.

\begin{acknowledgments}
One of the authors (M.H.L.) is grateful to Chong-Der Hu for valuable
discussions. We gratefully acknowledge financial support by the Republic of
China National Science Council Grant No. 95-2112-M-002-044-MY3.
\end{acknowledgments}

\end{document}